\documentclass[aps,prl,twocolumn,superscriptaddress,floatfix]{revtex4}

\usepackage{amsmath}
\usepackage{graphicx}
\usepackage{dcolumn}
\newcommand{\mypath}[1]{./#1}

\begin{document}

\title{Unveiling First Order CMR Transitions in the Two-Orbital Model for Manganites}

\author{Cengiz \c{S}en}
\affiliation{Department of Physics and Astronomy, The University of Tennessee, Knoxville, TN 37996}
\affiliation{Materials Science and Technology Division, Oak Ridge National Laboratory, Oak Ridge, TN 32831}

\author{Gonzalo Alvarez}
\affiliation{Computer Science \& Mathematics Division
and Center for Nanophase Materials Sciences, 
Oak Ridge National Laboratory, Oak Ridge, TN 37831}

\author{Elbio Dagotto}
\affiliation{Department of Physics and Astronomy, The University of Tennessee, Knoxville, TN 37996}
\affiliation{Materials Science and Technology Division, Oak Ridge National Laboratory, Oak Ridge, TN 32831}

\date{\today}

\begin{abstract}
Large scale Monte Carlo simulation results for the two-orbital model for manganites, including
Jahn-Teller lattice distortions, are here presented. At hole density $x=1/4$ and in the vicinity of 
the region of competition between the ferromagnetic metallic and spin-charge-orbital ordered 
insulating phases, the colossal magnetoresistance (CMR) phenomenon is observed with a magnetoresistance
ratio $\sim 10,000\%$. Our main result is that this CMR transition is found to be of first order in some
portions of the phase diagram,
in agreement with early results from neutron scattering, specific heat, and magnetization, 
thus solving a notorious
discrepancy between experiments and previous theoretical studies. The first-order characteristics of the transition survive, and
are actually enhanced, when weak quenched disorder is introduced.

\end{abstract}


\maketitle

{\it Introduction.}
The manganese oxides known as manganites continue attracting considerable attention
due to the presence of several competing tendencies in their rich phase
diagrams, with a variety of spin, charge, and orbital orders~\cite{otherreviews,review}. 
Moreover, these compounds display the famous CMR effect, where the resistivity is
drastically reduced by fields of a few Teslas. Early
theoretical investigations~\cite{science} remarked the importance of phase competition to understand
the CMR, which occurs when an insulating state, typically also spin-charge-orbital
ordered, is close in energy to the low-energy ferromagnetic (FM) metallic ground state
induced by double exchange. These effects were clear in simplified phenomenological models and
resistor-network simulations when in the presence of disorder~\cite{burgy}.
However, it is important to verify at a more fundamental level 
if the basic model Hamiltonians that are widely perceived as being
realistic for manganites, including double-exchange, superexchange, and lattice distortion tendencies,
are indeed compatible with the CMR phenomenology. Recently, our group and others initiated this effort
based on Monte Carlo (MC) 
simulations~\cite{verges,kumar,sen06,sen07}. This is a challenging task since
computer efforts based on the exact solution of the fermionic $e_{\rm g}$ sector, for 
fixed MC-generated classical $t_{\rm 2g}$ spin and lattice
configurations, have a CPU time growing as $N^4$, with $N$ 
the number of cluster sites times the number of
orbitals. For this reason, recent efforts at 
realistic electronic densities have used only one $e_{\rm g}$ orbital, considerably reducing
the CPU demands~\cite{sen07}. 
However, it is clear that the more complete 
two $e_{\rm g}$ orbitals model must be investigated for a full understanding of the CMR physics.
Alas, since this is far more time consuming, MC results for two orbitals are only 
scattered in the literature
and they have not reached sufficient accuracy to unveil the true
properties of this model in the CMR regime.

In addition, an important qualitative difference
still persists with regards to the {\it order} of the CMR transitions. 
Until now, MC simulations carried out
at CMR realistic hole densities, such as $x=0.33$ or $0.25$,
have indicated the presence of a rapid crossover, yet continuous,
transition from the low-$T$ FM metal to the high-$T$ paramagnetic 
(PM) insulator. This occurs
even in the clean limit of the MC simulations, i.e. without 
quenched disorder.
While these results are in agreement with the phenomenology of many CMR manganites~\cite{review},
there are notorious examples where the CMR transition is 
of first order, such as for
${\rm Pr_{0.55} (Ca_{0.75}Sr_{0.25})_{0.45} MnO_3}$~\cite{PCSMO}, 
${\rm Sm_{0.52}Sr_{0.48}MnO_3}$~\cite{SSMO,SSMO2}, 
${\rm (Nd_{0.5}Sm_{0.5})_{0.55} (Ba_{0.5}Sr_{0.5})_{0.45} MnO_3}$~\cite{NSBSMO}, 
and others. In fact, early experimental studies already reported
signs of irreversibility at the CMR transition 
for one of the most widely studied manganites, La$_{1-x}$Ca$_x$MnO$_3$ (LCMO)
at $x=0.33$, suggesting that its transition is weakly first-order~\cite{lynn96}. Subsequent 
magnetization and specific heat analysis clarified that indeed the $x=0.33$ LCMO transition
is of first order~\cite{mitchell}.
Recent investigations have also addressed the order of the CMR transition, revealing
multicritical characteristics in 
RE$_{0.55}$Sr$_{0.45}$MnO$_3$ (RE = Rare-Earth)~\cite{multicritical}.

Our main goal in this publication is to solve this puzzling theory-experiment disagreement by 
carrying out a large-scale computational study of the two-orbital 
model for manganites at $x=1/4$. It will be shown that our
MC data unveils the presence of a
first-order CMR FM-PM transition in this realistic model and density.

{\it Model and techniques.} The two-orbital model Hamiltonian used here is widely considered the minimal model
for the proper description of the electronic properties of manganites and it has been extensively discussed
before~\cite{hotta2001}. For this
reason only a schematic description will be here provided. The model
is given by $H = H_{\rm DE}+H_{\rm JT}+H_{\rm AF}+H_{\rm Dis}$~\cite{hotta-coul}. At every Mn site it
contains $e_{\rm g}$ electrons and $t_{\rm 2g}$ localized spins. 
$H_{\rm DE}$ represents the nearest-neighbors double exchange (DE) $e_{\rm g}$ electron 
hopping at infinite Hund coupling. This term favors ferromagnetism
away from the electronic density $n = 1$.
$H_{\rm JT}$ is the coupling between fermions and oxygen breathing
and Jahn-Teller lattice distortions, with coupling strength $\lambda$.
The sum $H_{\rm DE}+H_{\rm JT}$ is Eq.(4) of Ref.~\onlinecite{sen06}, with the hopping 
along $x$ for orbital $a = x^2-y^2$, $t_{\rm aa}$, as the energy unit.
$H_{\rm AF} = J_{\rm AF} \sum_{\langle {\bf ij} \rangle}{{\bf S}_{\bf i}}\cdot{{\bf S}_{\bf j}}$ is the standard nearest-neighbors 
Heisenberg antiferromagnetic coupling among the (classical) $t_{\rm 2g}$ spins. 
Finally, $H_{\rm Dis} = \sum_{\bf i} \Delta_{\bf i} n_{\bf i}$ is the quenched disorder
term, with $\Delta_{\bf i} = \Delta \times r_{\bf i}$ ($\Delta$ = disorder strength; 
$r_{\bf i}$ = bimodal random number 1 or -1).

{\it Details of the computer simulation.} The methodology of our MC simulations, based on the fermionic sector
exact diagonalization described before, is standard and readers should consult \cite{review} for details.
However, the extensive characteristics of the present
MC simulations merits a detailed discussion. 
The procedure was the following. 
For the clean-limit results $\Delta = 0$, a random spin and lattice configuration was 
chosen at the highest 
studied temperature, $T=0.33$, to initiate the runs. After 10,000 MC steps, $T$ 
was reduced and another 10,000 steps were performed. This
``cooling down'' process continued, using a grid with 23 temperatures (lowest 1/300)
that was denser near the critical temperatures $T_C$. 
After this already demanding first step,
the cooling-down results (obtained  
by measuring during the last 5,000 MC steps per $T$)
revealed sharp, yet continuous, transitions 
at all couplings. However, 
at particular $T$s in the MC time evolution indications 
of insufficient convergence were found. 
Thus, next for each $T$ the results were further refined
using additional 10,000 MC steps for extra thermalization, followed by 100,000
MC steps for measurements. By monitoring
the results during the final MC evolution it was observed that this large effort
produces now fairly stable results, revealing the first-order 
transitions in the region of phase competition discussed below.

The clean-limit effort needed standard computer clusters 
with $\sim 100$ nodes. However, 
a similar procedure with quenched disorder, requiring 40 $r_{\bf i}$ disorder 
configurations, 6 disorder strengths 
($\Delta$), and 20 temperatures for each $J_{\rm AF}$, would have been 
impossible.  Thus, the results with disorder reported below 
were obtained using 
the UT-ORNL Kraken supercomputer (Cray XT5), where up 
to 5,000 processors were employed simultaneously for periods of 24 hours.
This amounts to a total computational effort 
in Figs.~3(c) and 3(d) (see below) of $\sim$250,000 hours
($\sim$30 years if ran serially).

\begin{figure}
\centerline{\includegraphics[clip,width=8cm, height=7cm]{\mypath{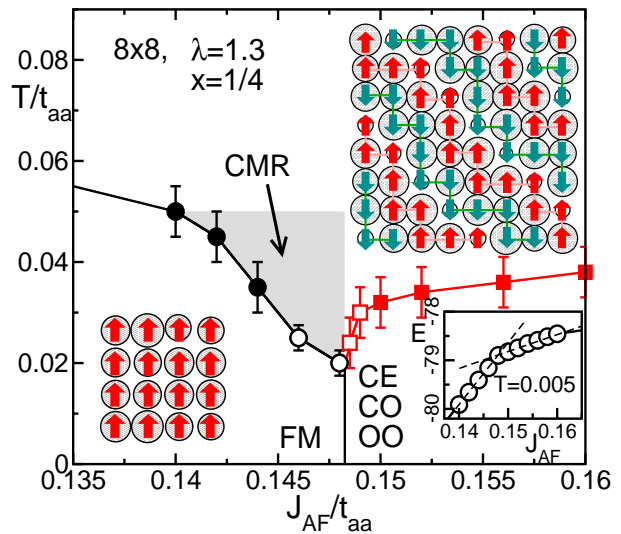}}}
\caption{(Color online) Clean-limit MC phase diagram of the $x = 1/4$ doped 
two-orbital model at $\lambda=1.3$ and on an $8\times8$ lattice. 
Critical temperatures are estimated
from spin structure factors (Fig.~2) and real-space spin-spin correlations (not shown). 
First (second) order transitions are indicated with open (filled) symbols, and the shaded region is where the
CMR is observed (Fig.~3).
Also shown are the $T$=0 arrangements of
classical spins and electronic charges (magnitude proportional to the radius of the circles) 
for the competing metallic and insulating states.
{\it Inset:} total energy vs. $J_{\rm{AF}}$. 
The straight lines crossing 
indicates a low-$T$ first-order transition. The spins were frozen to the FM and CE
perfect spin states and the oxygen distortions were MC relaxed.
}
\vspace{-0.6cm}
\label{Figure1}
\end{figure}

\begin{figure}
\flushleft{
\includegraphics[clip,width=8.0cm,height=8.0cm]{\mypath{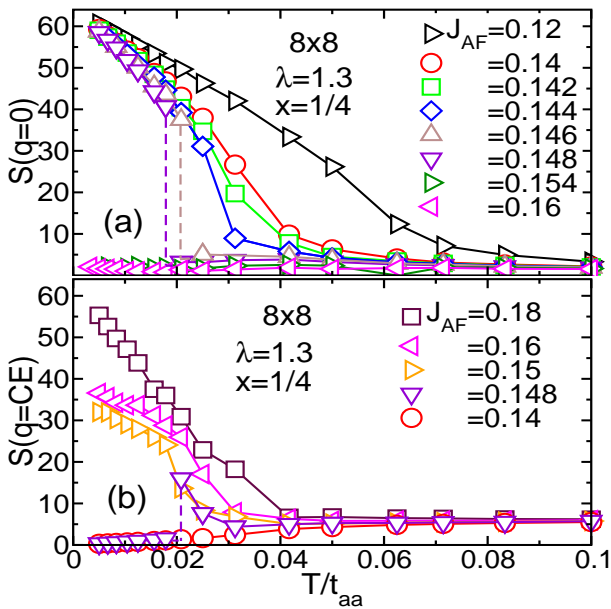}}}
\caption{(Color online) 
Spin structure factor $(S({\bf q}))$ vs. $T$ for the momenta corresponding to the 
(a) FM and (b) CE states. $S({\bf q})$ was used
to determine the critical temperatures in Fig~\ref{Figure1}. 
For the CE phase, the relevant momenta are  ${\bf q}$=($\pi/4,3\pi/4),~(\pi/2,\pi/2)$, 
and $(3\pi/4,\pi/4)$, as well as ${\bf q}^{\prime}$=${\bf q}$+$\pi$ (i.e. rotating
the zigzag chains by 90-degrees). 
}
\vspace{-0.6cm}
\label{Figure2}
\end{figure}

{\it Clean limit phase diagram.}
The clean-limit phase diagram of the two-orbital model 
is in Fig.~\ref{Figure1}.
The chosen coupling 
$\lambda = 1.3$ is representative of the regime where metallic
and Jahn-Teller distorted insulating phases are in competition. 
The hole density $x = 1/4$ is in the realistic range 
of experimental CMR investigations. Varying $J_{\rm AF}$, 
and via the  MC procedure described before,
a systematic analysis of spin and charge correlations
(e.g. 
see Fig.~\ref{Figure2}) leads to a phase diagram where the
FM metallic phase generated by the DE mechanism competes with
a previously discussed C$_{0.25}$E$_{0.75}$ insulator~\cite{hotta-CE} (shown in Fig.~\ref{Figure1}). 
This ``CE'' state is the
natural generalization to $x=1/4$ of the well-known
$x=1/2$ CE state~\cite{otherreviews,review,science}, and they only differ in the shape
of the zigzag chains. 
Like in all computational based efforts, here relatively
small clusters are used, but it is well known that locating the temperature
range where the relevant correlation lengths are as large as the 
cluster size provides qualitatively correct estimations of trends
and  critical temperatures.

The main novelty of the phase diagram in Fig.~\ref{Figure1}
is the identification of a first-order transition separating
the FM metallic phase from the PM high-$T$ state, in the 
coupling range close to the competing CE insulator. Previous MC investigations
had not reached sufficient accuracy to detect this first-order transition
at realistic hole densities, such as $x=1/4$. While previous efforts
had clearly established the first-order nature 
of the direct low-$T$ FM-CE transition
(see inset of Fig.~\ref{Figure1}), 
now observing the more subtle first-order FM-PM transition 
represents qualitative progress in the modeling of manganites.

Figure~\ref{Figure2}(a) illustrates how the spin structure factor varies with $T$. 
At {\bf q}=(0,0) and close to the region of CE competition, 
a discontinuous transition exists between the FM and PM states.  
There is also a first-order transition from the CE to the PM state, close to the FM region.
our new results reveal that the true phase diagram of clean-limit 
manganite models actually has ``multicritical''
characteristics, with the robust FM-CE first-order transition at low $T$ splitting
into still first-order FM-PM and CE-PM transitions with increasing $T$, each
ending at critical points (whose precise location is beyond our accuracy). 
These results are compatible with recent multicritical characteristics revealed in some 
manganites~\cite{multicritical}.

\begin{figure}
\centerline{
\includegraphics[clip,width=8.0cm]{\mypath{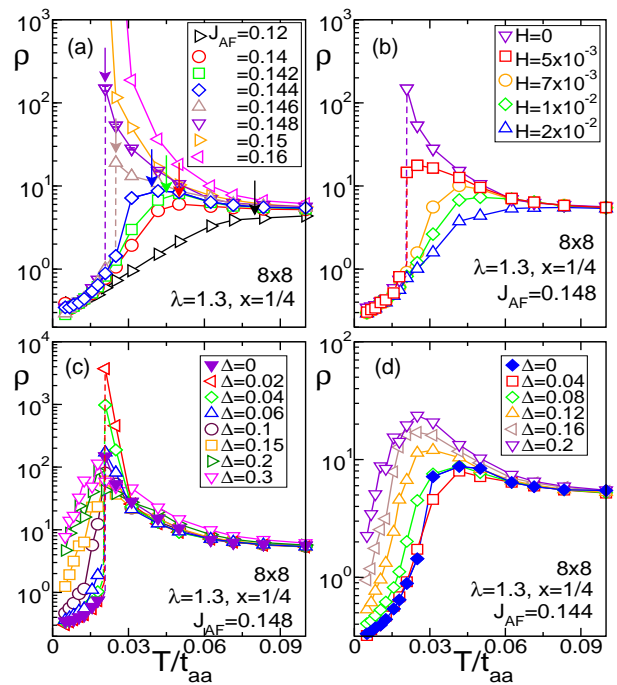}}}
\caption{(Color online) (a) $\rho$ vs. $T$, illustrating 
the metal-insulator transition varying $J_{\rm AF}$. Arrows indicate
the magnetic $T_C$s. Error bars are small (shown only at $J_{\rm AF}=0.148$).
(b) Magnetic field effects (along the $z$-direction) on the
$\rho$ vs. $T$ curves, showing the CMR effect. 
(c,d) Quenched disorder influence (averages over 40 disorder configurations) 
on the $\rho$ vs. $T$ curves 
for (c) $J_{\rm AF}$=$0.148$ and (d) $J_{\rm AF}$=$0.144$, with first-order and
continuous (but rapid) transitions, respectively. 
}
\vspace{-0.6cm}
\label{Figure3}
\end{figure}

\begin{figure}
\centerline{
\includegraphics[clip,width=8.0cm]{\mypath{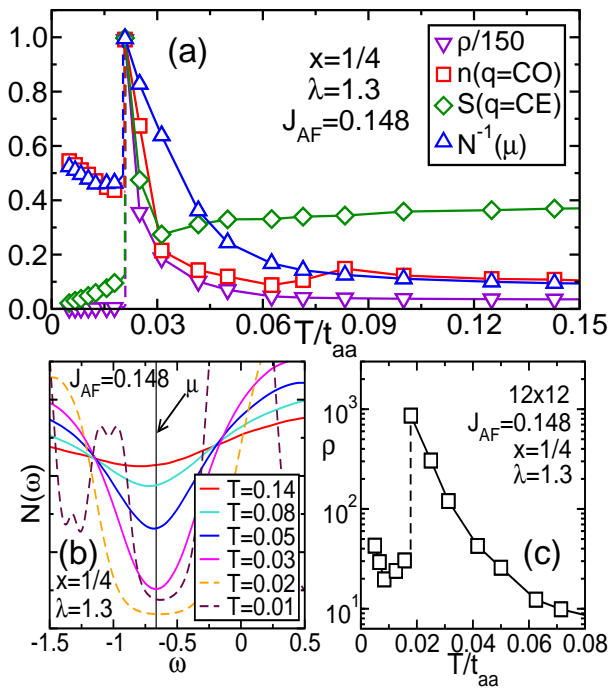}}}
\caption{(Color online)
(a) Spin $S({\bf q})$ and charge $n({\bf q})$ 
structure factors, shown together with the resistivity 
$\rho$  and the Fermi-level inverse density-of-states, vs. $T$. 
The ``q=CO" shorthand stands for the characteristic momenta of the CE state
($\bf{q}$=$(\pi/2,\pi/2)$,~$(\pi,\pi)$, and
$(3\pi/2,3\pi/2)$, and momenta obtaining by rotating the zigzag chains by
$\pi/2$). 
$n({\bf q})$ is obtained by Fourier-tranforming the quantum
correlations between local charge densities among $e_{\rm{g}}$ orbitals $x^2-y^2$ and $3z^2-r^2$, and averaging over the MC steps.
(b) Pseudogap in the density of states $N(\omega)$ as $T$ is reduced. (c) $\rho$ vs. $T$ for a 12$\times$12 cluster, showing also
a first-order transition.
}
\vspace{-0.4cm}
\label{Figure4}
\end{figure}

{\it CMR effect.} The first-order nature of the transitions observed here dramatically
affects the transport properties. 
Using the standard Kubo formula 
to calculate conductances~\cite{verges,sen06}, 
Fig.~\ref{Figure3} shows the resistivity ($\rho$) vs. $T$ obtained 
in our simulations.
Fig.~\ref{Figure3}(a) are results in the clean limit, 
varying $J_{\rm AF}$ in the region
of FM-CE competition. At couplings 
such as $J_{\rm AF}$=0.148, $\rho$ is
insulating upon cooling, closely following results for 
$J_{\rm AF}$=0.16 with a CE ground state.
However, at the FM transition $T_C$$\sim$0.02 ($\sim$ 100~K, if $t_{\rm aa}$=0.5~eV), $\rho$ becomes metallic via an abrupt discontinuity, 
in excellent agreement with several experiments~\cite{PCSMO,SSMO,SSMO2,NSBSMO,lynn96,mitchell}. As $T_C$ increases, moving the system further away from the CE state, the transition
becomes continuous and at $J_{\rm AF}$=0.12 the FM transition is 
barely noticeable in the slope of $\rho$ vs. $T$. For completeness, in Fig.~\ref{Figure4}(c) results using
a 12$\times$12 cluster for only one set of couplings $\lambda$-$J_{\rm AF}$ (due to its high CPU cost) are shown, 
indicating that cluster size effects are small~\cite{comment-cube}.

Figure~\ref{Figure3}(b) contains the $\rho$ curves in the presence of magnetic fields $H$.
The observed trends are again 
in excellent agreement with experiments~\cite{PCSMO,SSMO,SSMO2,NSBSMO,lynn96,mitchell}, with an 
overall rapid decrease of $\rho$ and with 
$\rho$-peak positions moving to higher $T$ with
increasing $H$, and all the curves merging at approximately room $T$ (i.e.  $T \sim0.06$ if $t_{\rm aa} \sim 0.5$ eV).
At a small field $H = 7\times10^{-3} t_{\rm aa}$, the magnetoresistance $(\rho(0)-\rho(H))/\rho(H)\times100\%$ is $\sim 10,000\%$
also in good agreement with CMR phenomenology.

{\it Influence of Quenched Disorder.} Experiments and theoretical calculations have shown
the importance that quenched disorder has over the CMR effect~\cite{review}. 
It is expected that the clean-limit fine tuning of couplings 
needed to obtain a CMR (Fig.~\ref{Figure1}) will be removed once
disorder is incorporated. Thus, it is important to analyze the influence of disorder on our results.
Using the on-site quenched disorder form described before, 
results are 
in Figs.~\ref{Figure3}(c,d). Panel (c) illustrates how the 
clean-limit first-order transition is eventually 
rendered continuous by increasing the
disorder strength.
However, the discontinuity in $\rho$ first increases
with increasing $\Delta$ before it is reduced. This 
result is compatible with the observed
multicritical behavior even with disorder~\cite{multicritical}. 
Also as in experiments, 
when the clean-limit transition is second order, 
quenched disorder decreases the $T$ 
where the peak occurs while increasing $\rho$ (panel (d)).

The CMR effect observed in our 
clean-limit 
MC simulations can be understood qualitatively via measurements of a 
variety of observables, similarly as in previous investigations~\cite{sen07}. 
For instance, Fig.~\ref{Figure4}(a) contains
the spin and charge structure factors, $S({\bf q})$ and $n({\bf q})$, at
the momenta of relevance for the CE phase, but in the CMR regime
where the $T=0$ ground state is FM. As $T$ decreases, not
only $\rho$ rapidly increases but so do $S({\bf q}_{CE})$ and $n({\bf q}_{CO})$, 
showing that the system behaves as if the
ground state were CE, developing robust CE short-range correlations.
However, at $T_C$ an abrupt transition occurs to the true FM ground
state. This switch from CE-dominated to FM-dominated characteristics
with cooling may occur if the high-$T$ short-range-ordered CE state
has a high entropy. Note also that at low-$T$, the charge correlations
in the FM state are still robust at short distances. Finally, also note
that at a $T^*/t_{\rm aa} \sim 0.07-0.08$, 
considerably higher than $T_C$, the CE tendencies
start developing upon cooling~\cite{argyriou}, and this occurs concomitantly 
with the presence of a density-of-states 
pseudogap (Fig.~\ref{Figure4}(b)), as also observed
in photoemission experiments~\cite{PG}. It is gratifying to observe similar
results above $T_C$ for both the
present model and the model studied in Ref.~\onlinecite{sen07}.
Finally, note that the many observations of FM signals above $T_C$ in
several manganites~\cite{review} are not incompatible with the clean-limit Fig.~\ref{Figure4} 
since quenched disorder is known to increase the strength of the FM component.

{\it Conclusions.}  The observation
of first-order CMR transitions in models for manganites was here reported,
solving a notorious theory-experiment discrepancy.
A large-scale computer simulation was needed to reach our conclusions.
Robust CMR ratios were found, as well as a CMR state above $T_C$ with short
range CE characteristics. Weak quenched disorder preserves 
the first-order transitions, as in recent experiments~\cite{multicritical}.

{\it Acknowledgment.} This work was supported by the NSF 
grant DMR-0706020 and by the Division of
Materials Sciences and Engineering, 
Office of Basic Energy 
Sciences, U.S. DOE. The computer simulations were possible in part 
by a NSF allocation of advanced computing resources at the Kraken (Cray XT5)
supercomputer located at the National Institute for Computational 
Sciences (\url{http://www.nics.tennessee.edu/}). 
A portion of this research was conducted at the Center for
Nanophase Materials Sciences at ORNL. 
This research used the SPF software (\url{http://www.ornl.gov/~gz1/spf/}).


\begin{thebibliography}{00}


\bibitem{otherreviews} 
M. Salamon and M. Jaime, Rev. Mod. Phys. {\bf 73}, 583 (2001);
J. De Teresa {\it et al.}, Nature {\bf 386}, 256 (1997);
M. Uehara {\it et al.}, Nature {\bf 399}, 560 (1999).

\bibitem{review} E. Dagotto {\it et al.}, Phys. Rep. {\bf 344}, 1 (2001).


\bibitem{science} A. Moreo {\it et al.}, Science {\bf 283}, 2034 (1999).

\bibitem{burgy} J. Burgy {\it et al.}, 
Phys. Rev. Lett. {\bf 87}, 277202 (2001);
{\it ibid}, Phys. Rev. Lett. {\bf 92}, 097202 (2004); M. Mayr {\it et al.}, Phys. Rev. Lett. {\bf 86}, 135 (2001).

\bibitem{verges} J. A. Verg\'es {\it et al.}, Phys. Rev. Lett. {\bf 88}, 136401 (2002).

\bibitem{kumar} S. Kumar and P. Majumdar, Phys. Rev. Lett. {\bf 96}, 016602 (2006); {\it ibid},
Phys. Rev. Lett. {\bf 96}, 136601 (2005).

\bibitem{sen06} C. \c{S}en {\it et al.}, Phys. Rev. B {\bf 73}, 224441 (2006).

\bibitem{sen07} C. \c{S}en {\it et al.}, Phys. Rev. Lett. {\bf 98}, 127202 (2007).

\bibitem{PCSMO} Y. Tomioka and Y. Tokura, Phys. Rev. B {\bf 66}, 104416 (2002). 

\bibitem{SSMO} Y. Tomioka {\it et al.},
Phys. Rev. B {\bf 74}, 104420 (2006).

\bibitem{SSMO2} Y. Tomioka {\it et al.},
Phys. Rev. B {\bf 80}, 174414 (2009). 



\bibitem{NSBSMO} Y. Tomioka and Y. Tokura, Phys. Rev. B {\bf 70}, 014432 (2004).

\bibitem{lynn96} J. W. Lynn {\it et al.}, 
Phys. Rev. Lett. {\bf 76}, 4046 (1996).

\bibitem{mitchell} D. Kim {\it et al.}, 
Phys. Rev. Lett. {\bf 89}, 227202 (2002).

\bibitem{multicritical} L. Demko {\it et al.}, 
Phys. Rev. Lett. {\bf 101}, 037206 (2008).


\bibitem{hotta2001} T. Hotta {\it et al.}, 
Phys. Rev. Lett. {\bf 86}, 4922 (2001).


\bibitem{hotta-coul} The Hubbard $U$ effects
are mild at large Hund coupling,
T. Hotta {\it et al.}, Phys. Rev. B {\bf 62}, 9432 (2000).

\bibitem{hotta-CE} T. Hotta {\it et al.}, Phys. Rev. Lett. {\bf 90}, 247203 (2003).

\bibitem{comment-cube} Preliminary results
on 4$\times$4$\times$4 clusters (not shown) suggest that a similar CMR peak 
is obtained in three dimensions, but the first order transition is smeared 
by the {\it frustration} effect caused by the geometry of the CE $x=1/4$ zigzag chains 
that do not fit into 4$\times$4 layers. Simulations 
on unfrustrated 8$\times$8$\times$8 clusters are currently impossible, justifying
why our present effort has focused on two dimensions. Fortunately, there are no 
reasons to suspect that two and three dimensions will behave differently with
regards to the first-order transition.


\bibitem{argyriou} D. N. Argyriou {\it et al.}, Phys. Rev. Lett. {\bf 89}, 036401 (2002).

\bibitem{PG} D. Dessau {\it et al.}, Phys. Rev. Lett. {\bf 81}, 192 (1998);
A. Moreo {\it et al.}, Phys. Rev. Lett. {\bf 83}, 2773 (1999).


\end{thebibliography}
\end{document}